\documentclass{article}

\usepackage{arxiv}
\usepackage{lipsum} 
\usepackage[version=4]{mhchem}
\usepackage{siunitx}
\DeclareSIUnit\Molar{M}
\usepackage{graphicx}
\usepackage{hyperref}
\usepackage{amsfonts}
\usepackage{amsbsy}
\usepackage{amssymb}
\usepackage{xcolor}
\usepackage[normalem]{ulem}
\usepackage{float}
\usepackage{multirow}
\usepackage{ragged2e}
\usepackage[utf8]{inputenc} 
\usepackage[T1]{fontenc}    
\usepackage{hyperref}       
\usepackage{url}            
\usepackage{booktabs}       
\usepackage{amsfonts}       
\usepackage{nicefrac}       
\usepackage{microtype}      
\makeatletter
\def\blfootnote{\xdef\@thefnmark{}\@footnotetext}
\makeatother
\usepackage{graphicx}
\graphicspath{ {./images/} }

\title{Identifying Attention-Deficit/Hyperactivity Disorder through the electroencephalogram complexity}

\author{
   \noindent Dimitri Marques Abramov\textit{$^{a}$},  Henrique Santos Lima \textit{$^{b}$},  Vladimir Lazarev\textit{$^{a}$},\\ \textbf{Paulo Ricardo Galhanone,\textit{$^{a}$} and Constantino Tsallis\textit{$^c$}}}

\begin{document}
\maketitle

\begin{abstract}
There are reasons to suggest that a number of mental disorders may be related to alteration in the neural complexity (NC). Thus, quantitative analysis of NC could be helpful in classifying mental and understanding conditions. Here, focusing on a methodological procedure, we have worked with young individuals, typical and with attention-deficit/hyperactivity disorder (ADHD) whose NC was assessed using q-statistics applied to the electroencephalogram (EEG). The EEG was recorded while subjects performed the visual Attention Network Test (ANT) and during a short pretask period of resting state. Time intervals of the EEG amplitudes that passed a threshold were collected from task and pretask signals from each subject. The data were satisfactorily fitted with a stretched $q$-exponential including a power-law prefactor(characterized by the exponent c), thus determining the best $(c, q)$ for each subject, indicative of their individual complexity. We found larger values of $q$ and $c$ in ADHD subjects as compared with the typical subjects both at task and pretask periods, the task values for both groups being larger than at rest. The $c$ parameter was highly specific in relation to DSM diagnosis for inattention, where well-defined clusters were observed. The parameter values were organized in four well-defined clusters in $(c, q)$-space. As expected, the tasks apparently induced greater complexity in neural functional states with likely greater amount of internal information processing. The results suggest that complexity is higher in ADHD subjects than in typical pairs. The distribution of values in the $(c, q)$-space derived from $q$-statistics seems to be a promising biomarker for ADHD diagnosis.
\end{abstract}

\blfootnote{\textit{$^{a}$~Instituto Nacional da Saúde da Crianca, da Mulher e do Adolescente Fernandes Figueira. Fundação Oswaldo Cruz, Avenida Rui Barbosa 716, Flamengo, Rio de Janeiro 22250-020, Brazil.\\ 
E-mail: dimitri.abramov@iff.fiocruz.br }}

\blfootnote{\textit{$^{b}$~Centro Brasileiro de Pesquisas Fisicas, Rua Xavier Sigaud 150, Rio de Janeiro-RJ 22290-180, Brazil.\\ 
E-mail: hslima94@cbpf.br }}

\blfootnote{\textit{$^{c}$~Centro Brasileiro de Pesquisas Fisicas and National Institute of Science and Technology of Complex Systems, Rua Xavier Sigaud 150, Rio de Janeiro-RJ 22290-180, Brazil \\
Santa Fe Institute, 1399 Hyde Park Road, Santa Fe, 
 New Mexico 87501, USA \\
Complexity Science Hub Vienna, Josefst\"adter Strasse 
 39, 1080 Vienna, Austria \\
E-mail: tsallis@cbpf.br}}

\section*{Introduction}

Complexity is a property of many systems in the Universe related to integration among system's components by long-range correlations in a multiscale organization. As a result, these systems are non-reducible in their constituents, capable of transitioning between different states or configurations~\cite{Tsallis2009, BassetGazzaniga2011}. 

The lifeforms are dissipative systems that manifest endothermic long-term stability due to their capacity for self-organization since they are complex, controlling the entropy through allostatic mechanisms, which keep and regulate homeostasis in the face of environmental stressors \cite{Kelso1995, Baffy2014}. Therefore, pathological states, typically non-adaptive, would likely result in a decrease in the organism's complexity \cite{Peng1993, Goldberger1997, Keller2020}. In recent decades, several studies have systematically identified alterations in the brain complexity of individuals with mental disorders. \cite{YangTsai2013, Khesmiri2020, Lauetal2022}. However, the meaning of those changes in complexity remains elusive. Indeed, some of those studies have indicated higher complexity in patients as compared to typical subjects.

Many procedures based on different theoretical paradigms have been developed to quantify complexity in the human brain activity, mainly based on classical methods such as approximate entropy or Lempel–Ziv complexity, using indices of predictability and regularity of time series from the electroencephalogram (EEG) \cite{Lauetal2022}. In 1988, a generalization of the standard Boltzmann-Gibbs statistical mechanics (BGSM) --- non-extensive statistical mechanics (NESM), known as $q$-statistics --- was introduced \cite{Tsallis1988}. The NESM functions  satisfactorily describe the behavior of wide classes of natural, social, and artificial systems \cite{Tsallis2009,Ludescher2011, Bogachev2014,Greco2020}. Recently, we have demonstrated the applicability of $q$-statistics to describe brain complexity by analyzing the temporal regularity of the EEG signal \cite{Nosso2023}, fitting a $q$-exponential function upon the empirical distribution of the probabilities $y$ related to the occurrence of the events $x$ (in this case, events are time distances between EEG amplitudes that passes down a threshold), as follows:
\begin{equation}
y_q=a_q\,x^{c_q}/[1+(q-1)b_q \,x^{h_q}]^{\frac{1}{q-1}}\,
\label{qstr}
\end{equation}
where $a_q$ is the normalization constant, and $(b_q, c_q,h_q)$ are positive parameters.  The parameter  $c_q$ denotes the slope of the left tail, while the index  $q$ and the parameter $h_q$ have a straightforward relation with the slope of the right tail in the $q$-stretched exponential function. Thus, $q$ departs further from unity as the range of the spatial-temporal correlations increases; in the limit of short-range correlations, we return to the original BG exponential ($q=1$), which corresponds to strong chaos. In turn, the index $c_q$ characterizes the number of degrees of freedom. Physically, this parameter is related to the degeneracies of the physical states \cite{Reichlbook}. Some other complex systems are modeled by similar $q$-exponential functions with a power-law function, ($x^c_q$), such as the financial market \cite{TsallisAnteneodoBorlandOsorio2003}, air traffic \cite{Mitsokapasetal2021}, or COVID-19 spreading \cite{TsallisUgur2020}.

To study brain complexity and its relation to mind disorders, we believe that $q$-statistics can be an effective and consistent approach due to its wide applicability in describing the complexity of a wide range of systems and its simplicity in implementation. 

Here, we have evaluated the neural complexity in the EEG of typical boys and those with diagnostics of attention-deficit/hyperactivity disorder (ADHD) by $q$-statistics. Although ADHD is a condition with relatively recognized biological bases, accurate biomarkers for precise diagnosis and understanding of its mechanisms remain undefined \cite{Buitelaar2022}.  Some previous studies on NC were discordant regarding whether the EEG complexity in individuals with ADHD was greater or lesser than that of their typical peers using, as measuring methods,  Multiscale entropy, Fuzzy entropy, Lempel-Ziv complexity, among others~\cite{Lauetal2022,Ruiz2023,Hernandez2023}. However, some findings have shown that NC is higher in children with ADHD compared to their peers and ADHD adults \cite{Hernandez2023}. This discrepancy may be influenced by the presence of various mental comorbidities that adults with ADHD commonly manifest.  Here, we have described neural complexity from the EEG of the same ADHD and typical subjects that we had previously studied using other psychophysiological methods. We observed marked differences of event-related potentials \cite{Abramov2019a} and EEG topography \cite{Lazarev2016} between the groups, as well as high accuracy with respect to DSM diagnostics criteria by multivariate analysis \cite{Abramov2019b}. 

By analyzing EEG complexity through $q$-statistics, we aim to show relevant differences between the typical and ADHD boys. Let us anticipate that $q$-statistics will effectively cluster the subjects based on the $q$ and $c$ parameters, which are uncorrelated quantities.

\section*{Methods}

\subsection*{Subjects and procedures}
We examined 19 typical and 19 ADHD boys, aged from 11 to 13 years, performing the Posner's Attention Network Test (ANT). The EEG's of the subjects were recorded using a Nihon Kohden NK1200 EEG System at 20 scalp points according to the International 10/20 System, with linked biauricular reference (A1+A2) at a sampling rate of 1000 Hz. The visual task required high cognitive effort during testing Posner's alertness, spatial orientation, and executive dimensions of sustained attention. The test lasted 20 minutes. For details, see Kratz et al. \cite{Kratz2011} and Abramov et al. \cite{Abramov2019a}. Before the ANT task began, a short segment of EEG lasting about 5 minutes was collected (pretask), with open and closed eyes.

Two physicians evaluated the subjects independently of each other in order to reduce the bias associated with interpretation subjectivity. The classification of subjects regarding ADHD was based on the DSM-IV-TR criteria (at least six symptoms of inattention or hyperactivity/impulsivity were to be  satisfied). The children had to also present mental/behavioral dysfunctionality in at least two environments (e.g., school and home). The criteria for inattention and hyperactivity/impulsivity symptoms were assessed through direct questioning of the parents regarding their child's characteristics, with responses recorded as "yes" or "no" for each criterion. A score between 0 and 18 was computed based on these responses. The children in this study were free of pharmacological treatments and received psychopedagogical support for school activities. Boys with neurological or psychiatric comorbidities were excluded from the study. IQ was assessed using a reduced version of the Wechsler Intelligence Scale for Children (WISC) test, with the Vocabulary and Block Design subtests \cite{Mello2011}.

All subjects and their parents gave us written informed consent to participate in this study, which was performed following all international and local ethical rules after being approved by our independent ethical board (CEP-IFF), registered under CAAE 08340212.5.0000.5269 (2013).  

\subsection*{Data Analysis}
The EEG signal was filtered using a convolution approach with a vector of size $t$ ( 1 bin = 1 milisecond) defined by:
\begin{equation} 
V = exp(-x^2/2) 
\end{equation} 
where $x$ is ranging from $-\sqrt{2}$ to $\sqrt{2}$ with step size $t$. For low-pass filtering, $t$ = 10 ms (100\,Hz). And for high pass filtering, the output from convolution of $V$ with $t$ = 2000 \,ms (0.5\,Hz) was subtracted from the original signal. Baseline slow oscillations as well the muscle artifacts were suitably minimized. For 60\,Hz suppression (a very deterministic artifact from power net), we adopted band-pass filtering using a Fast Fourier Transform approach. No other signal handling was done. 

The negative part of the signal was used to perform the regularity analysis, where blink artifacts had little effect on the signal after filtering. We truncated the amplitudes smaller than -200 $\mu$V  (with higher absolute value) because no brain sources physiologically generate larger ones in the EEG. From the remaining negative part of the signal, we calculate the standard deviation and every amplitude that passes down the threshold of -1.0 std. dev. was considered as a time event (see \cite{Nosso2023} for more detailed explanation). The histogram of these intervals was computed within 1000 classes from each of the 20 EEG channels. The classes were nearly 2 ms long, but EEG signals with several large amplitude artifacts (truncated at -200 microvolts) presented slightly higher thresholds, which resulted in distributions with larger time bins. These distributions of frequencies were normalized to probability distributions, the integral thus being equal to unity. 

The analysis of the normalized frequencies of the inter-occurrence times leads to a fitting by the function currently emerging in nonextensive statistical mechanics, given by ~\eqref{qstr}. The parameters of the function were empirically determined by fitting using the least squares method. However, convergence is carried out by estimation, given computational limitations that make it impossible to execute all loops within the appropriate slopes. For greater precision of the parameters c and q through a second fitting, we analytically determined the parameters b and h through the respective linear functions obtained from the correlation of these two parameters with $q$, originating from a first fitting (see Fig. \ref{six}). The straight lines for $\ln {b}$ and for $h$,  for typical and ADHD individuals, namely $\ln \, b_{typical},\,h_{typical}$ and $\ln\, b_{ADHD},\,h_{ADHD}$, are as follows:

\begin{equation}
\begin{array}{c}
\begin{array}{rl}
\ln{b}_{\text{typical}} & \approx \begin{cases} 
-10.12 \, q + 12.5 & \text{if } \text{Pretask} \\
-7.6 \, q + 7.53 & \text{if } \text{Task} 
\end{cases} \\
h_{\text{typical}} & \approx \begin{cases} 
2.72 \, q - 2.73 & \,\,\,\,\,\,\,\text{if } \text{Pretask} \\
2.23 \, q - 1.65 &\,\,\,\,\,\,\, \text{if } \text{Task} 
\end{cases} 
\end{array}
\\ \\
\begin{array}{rl}
\ln{b}_{\text{ADHD}} & \approx \begin{cases} 
-9.12 \, q + 10.61 & \text{if } \text{Pretask} \\
-9.64 \, q + 11.33 & \text{if } \text{Task} 
\end{cases} \\
h_{\text{ADHD}} & \approx \begin{cases} 
2.75 \, q - 2.57 & \,\,\,\,\,\,\,\text{if } \text{Pretask} \\
2.8 \, q - 2.55 & \,\,\,\,\,\,\,\text{if } \text{Task} 
\end{cases} 
\end{array}
\end{array}
\label{str}
\end{equation}

We bind the distributions of all channels of each subject for best convergence. The probability distribution must obey the normalization condition, i.e., the integral $\int_0^{\infty}x^{c-h/(q-1)}\,dx$ can not diverge. This implies $(c+1)(q-1)/h<1$. Our results satisfy this condition as shown in Fig. \ref{normcond}.

After fitting, individual values of $c$ and $q$ in each condition and group (as well averaged ones) were plotted on a ($q \times c$) space to observe sample dispersion and possibly clustering. The Mann-Whitney U-Test inferred differences between groups and conditions. Correlations between parameters and other variables (DSM scores, I.Q. etc) were explored using the Pearson or Spearman Tests.

\section*{Results}

 Since we had previously ascertained that typical human EEG distributions satisfy $c >$ 1 in function ~\eqref{qstr} (at least for mid-parietal channel) \cite{Nosso2023}, we adopted $c\le 1$ at least in one channel of any condition as an exclusion criterion for the subject. So, the distributions that we regarded were those exhibiting left and right tails, representing observations from both typical and ADHD subjects. The dataset with such distributions corresponded to 15 typical and 15 ADHD boys.

\begin{figure*}
\centering
\includegraphics[width=\textwidth]{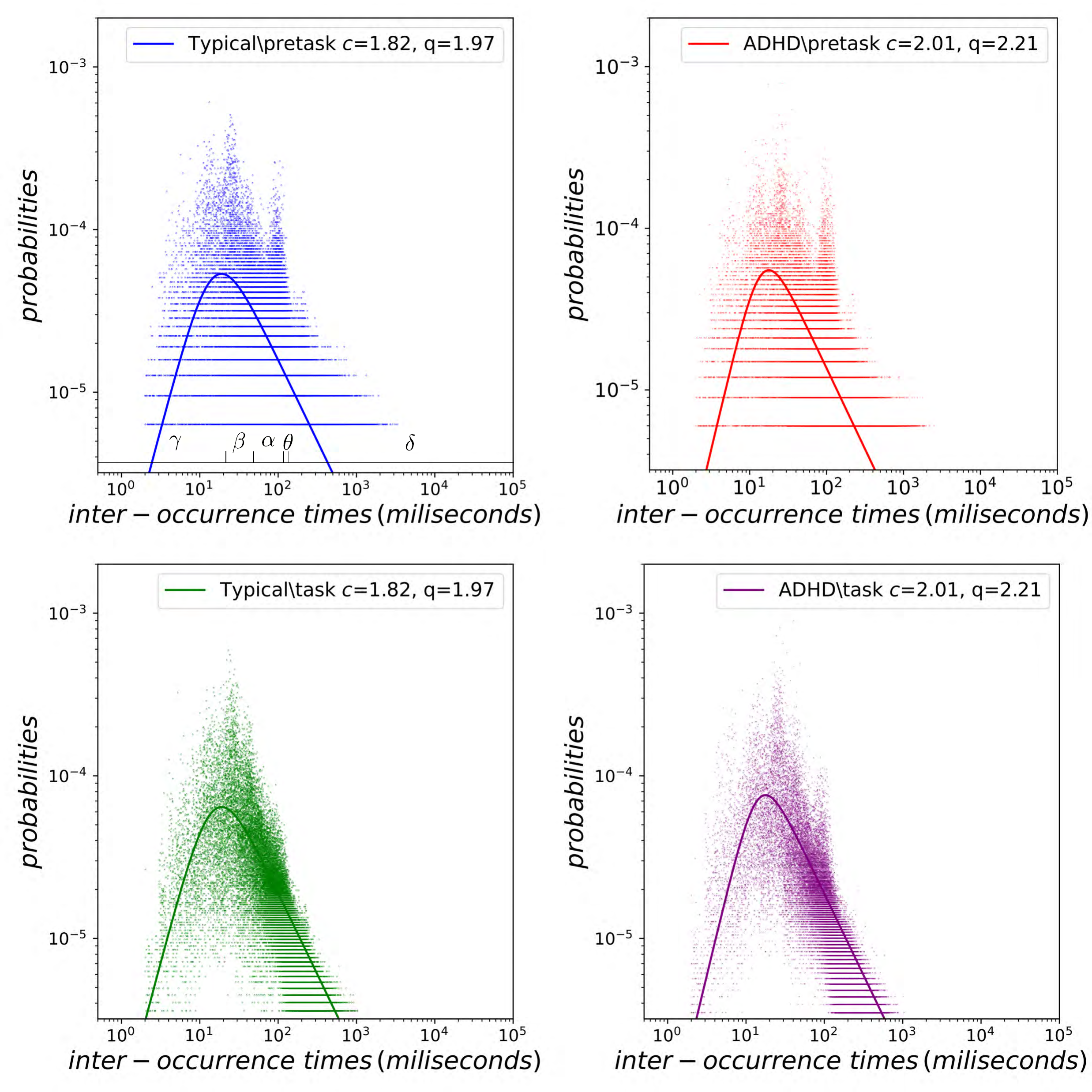}
\centering
\caption{ Probabilities versus inter-occurrence time events for  the datasets of all typical and ADHD subjects in pretask and task conditions and their corresponding fittings. (Top left) Typical\textbackslash pretask,(top right) ADHD\textbackslash pretask (bottom left) Typical\textbackslash task, and (bottom right) ADHD\textbackslash task. The Greek letters in 
 Typical\textbackslash pretask represent the following frequencies $\gamma$ (frequencies greater than 32 Hz), $\beta$ (between 13 and 31 Hz), $\alpha$ (between 8 and 12 Hz), $\theta$ (between 4 and 7 Hz), and $\delta$ (smaller than 4Hz), respectively. Notice that the data dispersion corresponding to pretask EEGs is larger than that corresponding to the task EEGs, possibly due to the time duration of the EEGs.}
 \label{fittingall}
\end{figure*}

\begin{figure*}
\centering
\includegraphics[width=6.4cm]{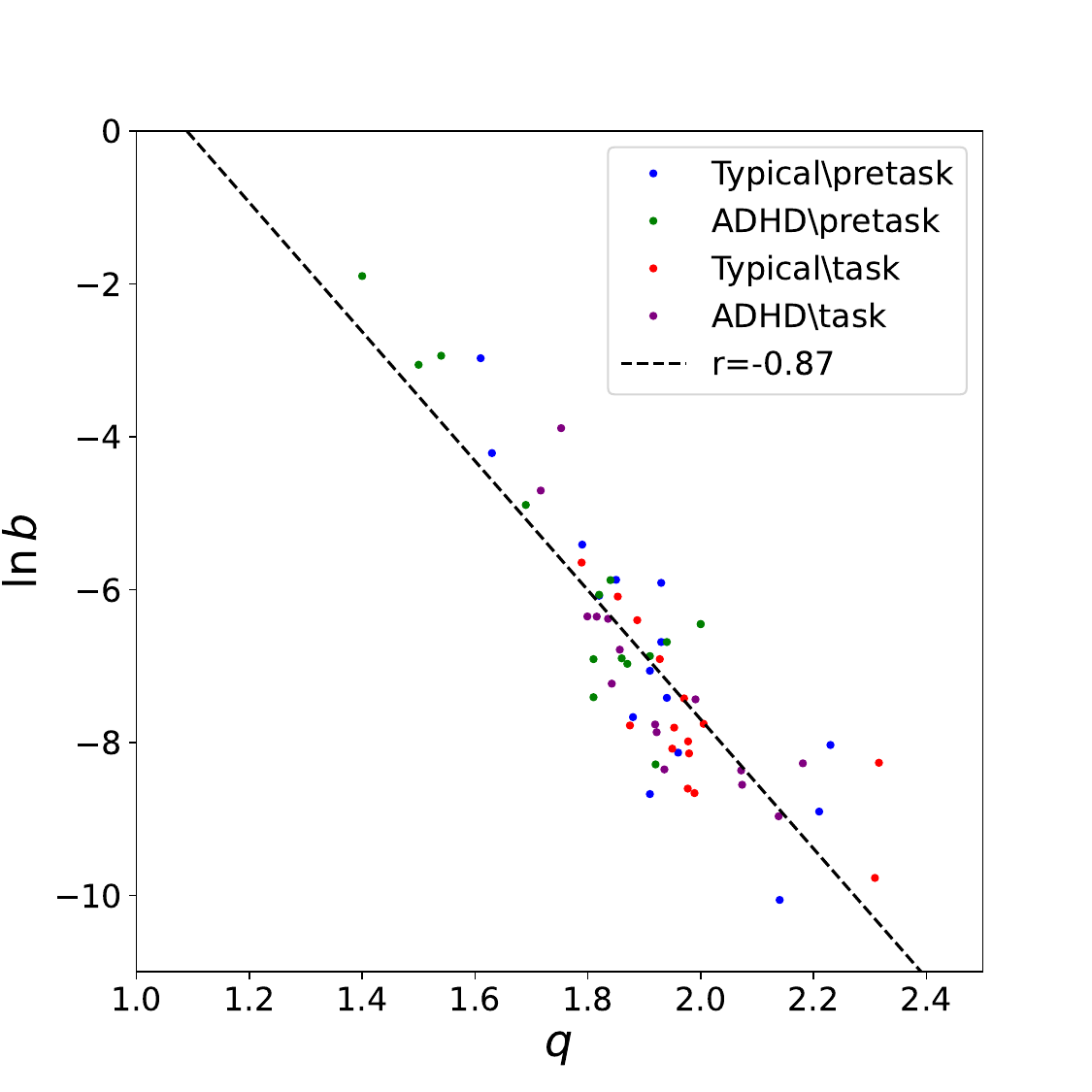}
\includegraphics[width=6.4cm]{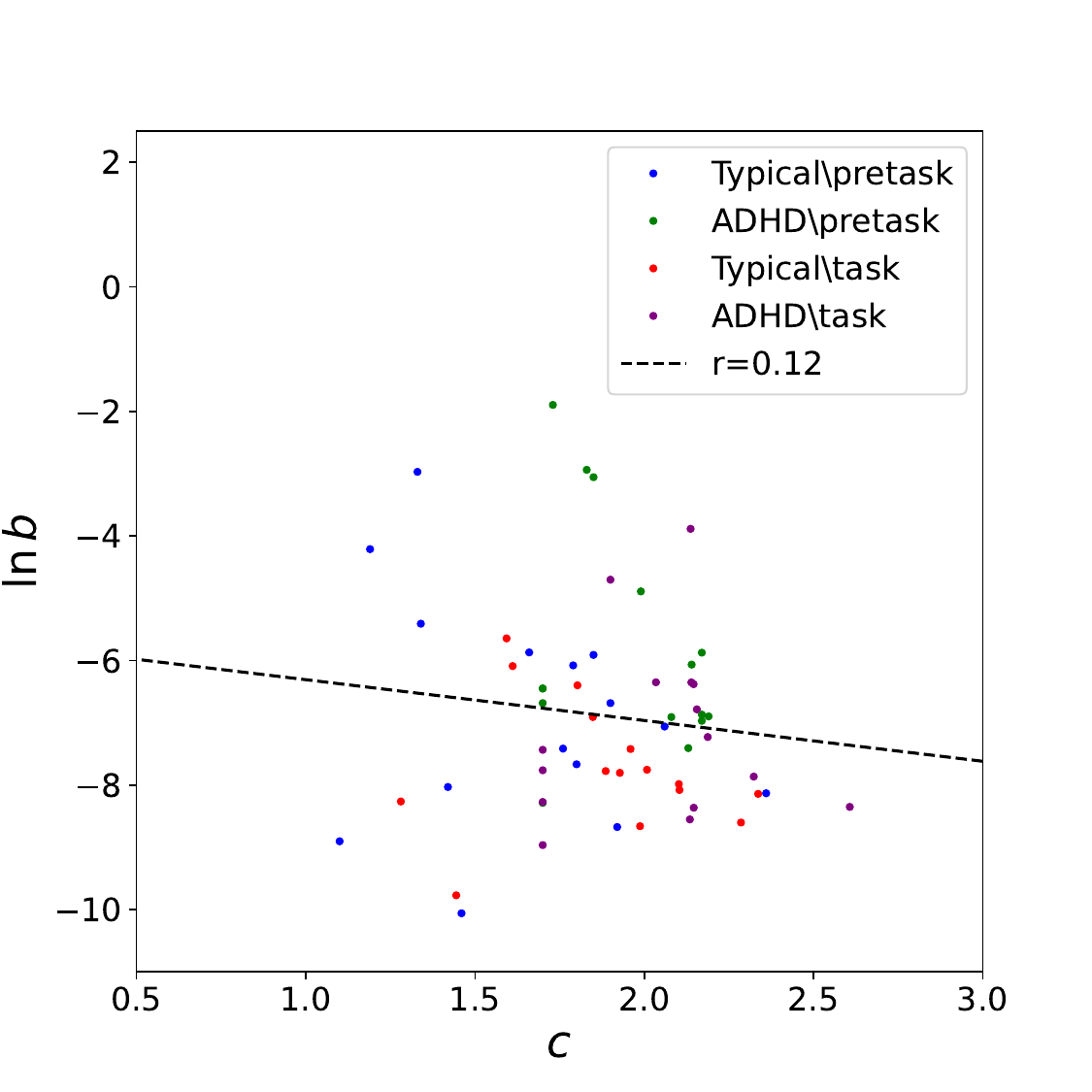}\\
\includegraphics[width=6.4cm]{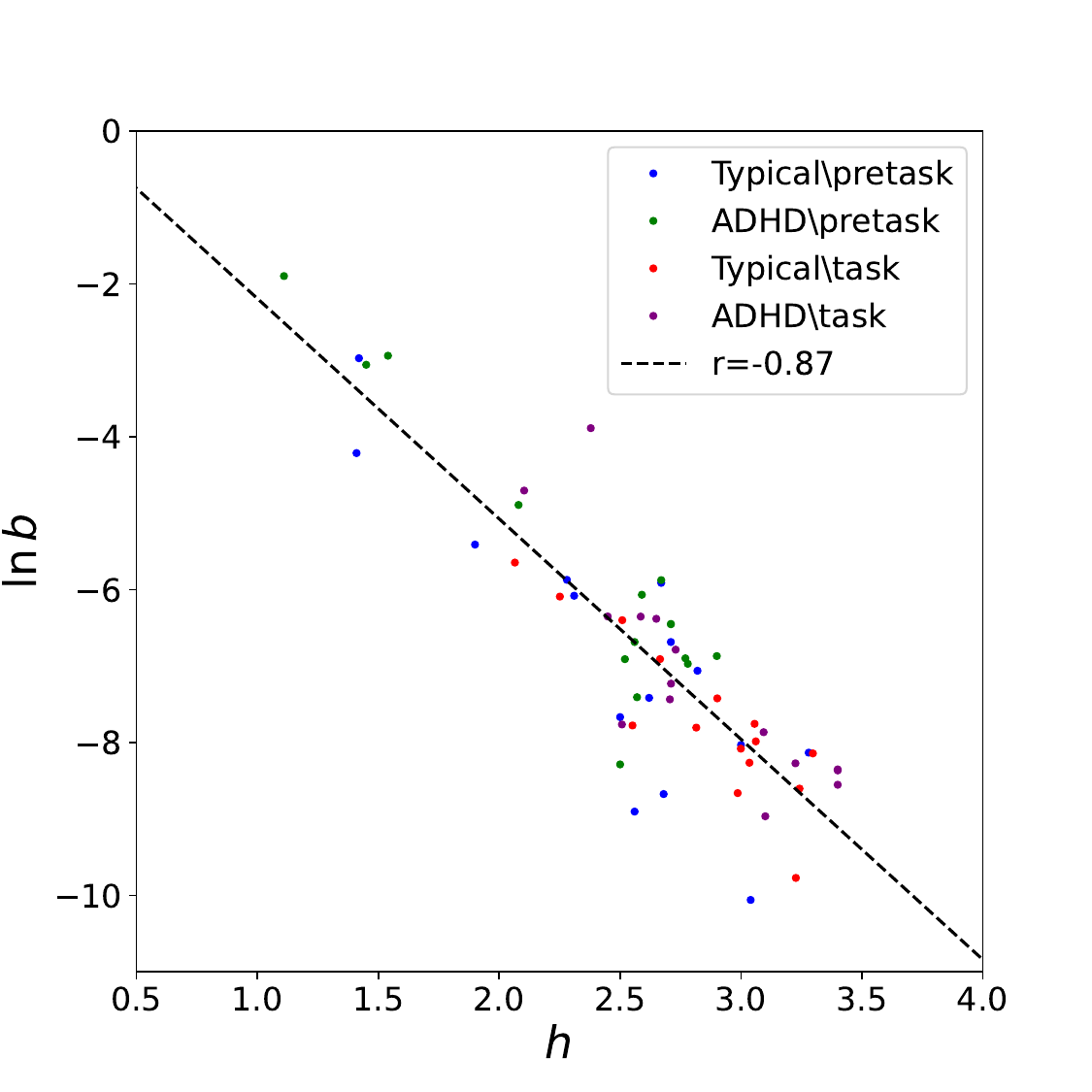}
\includegraphics[width=6.4cm]{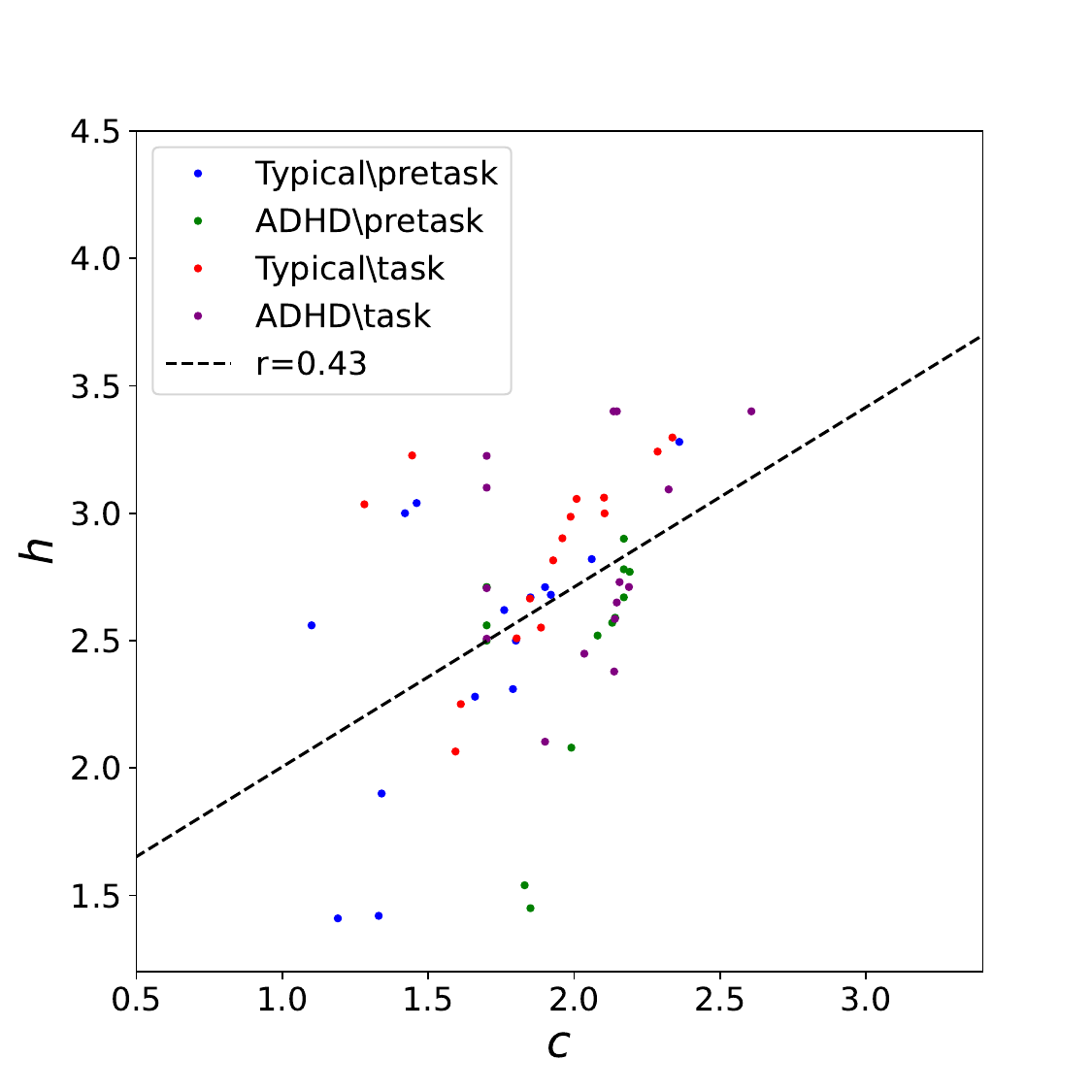}\\
\includegraphics[width=6.4cm]{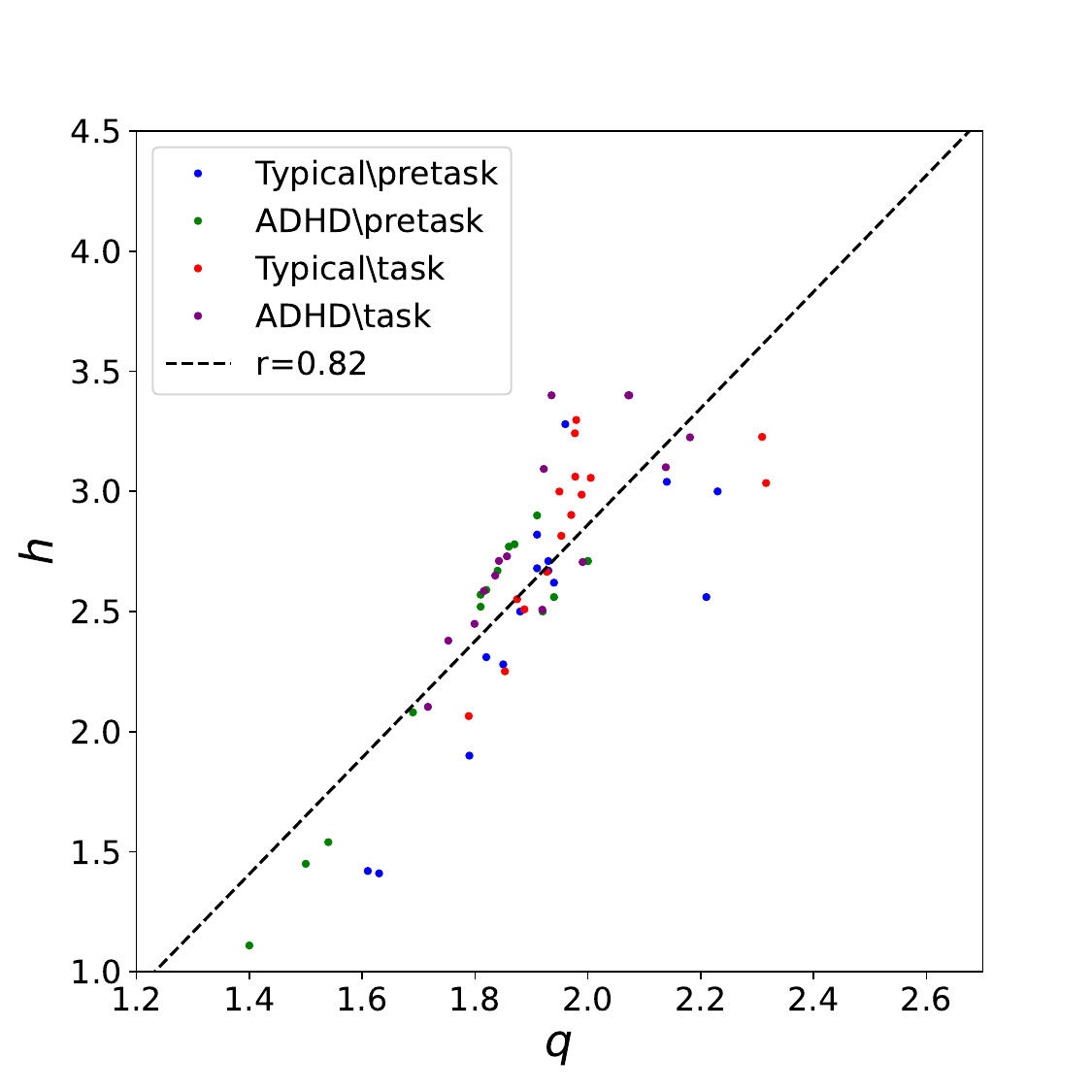}
\includegraphics[width=6.4cm]{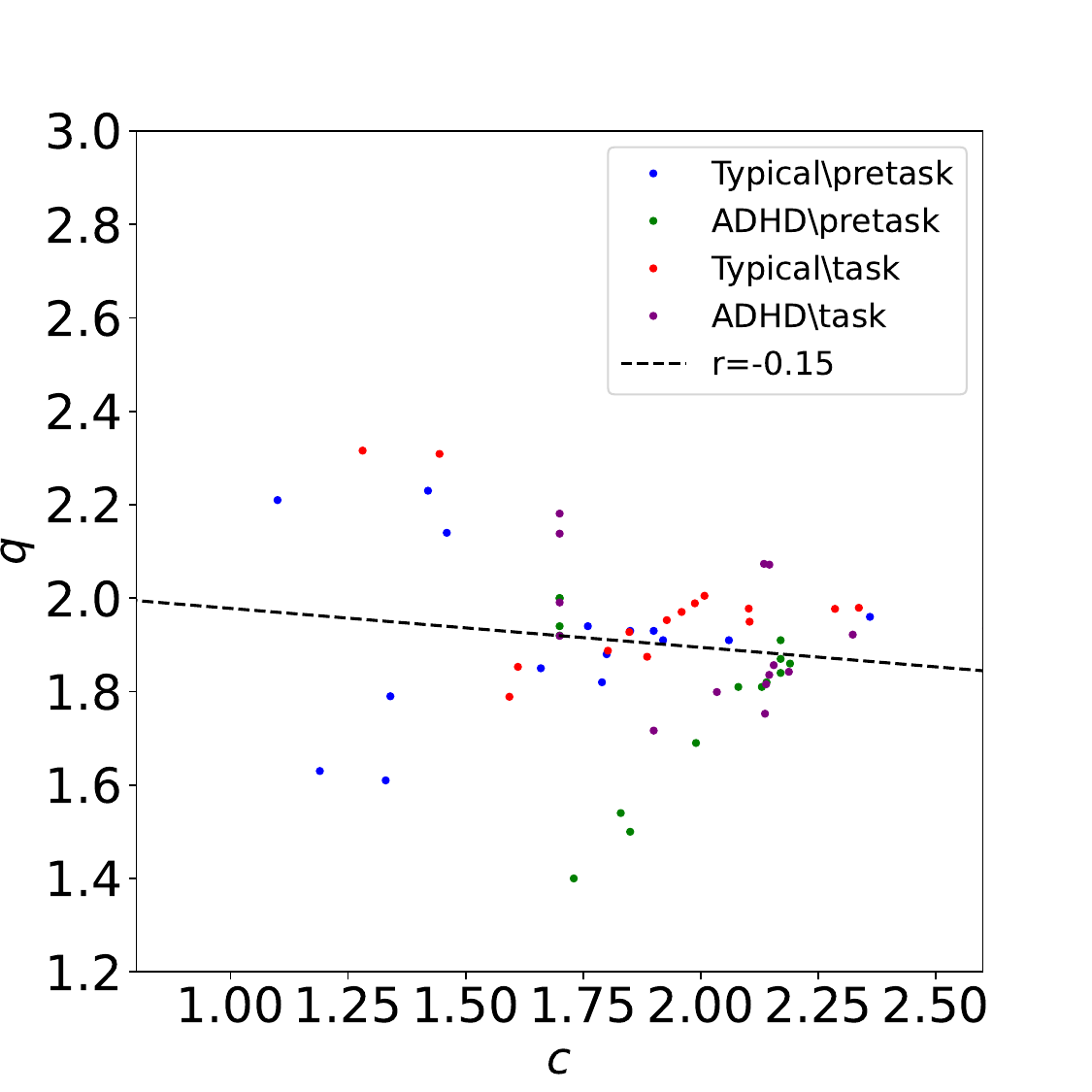}
\centering
\caption{(Top left) $\ln{b}\times q$, (top right) $\ln{b}\times c$, (middle left) $\ln{b}\times h$, (middle right) $h \times c$, (bottom left) $h \times q$, and (bottom right) $q \times c$. The r(Pearson) of the dashed straight lines in the same order are -0.87, 0.12, -0.87, 0.43, 0.82, and -0.15. The blue, green, red, and purple points represent typical\textbackslash pretask, ADHD\textbackslash pretask, typical\textbackslash task, and ADHD\textbackslash task, respectively.  }
\label{six}
\end{figure*}

\begin{table*}[htbp]

    \centering
    \caption{Relevant and confounding variables\\
    (*) SD = standard deviation;
    (**) in hours/week
    (***) Income per month in Brazilian minimum wages}
    \small
    \resizebox{12cm}{!}{
    \begin{tabular}{lcccccc}
        \toprule
        Variable & Typical(mean) & Typical(SD*) & ADHD(mean) & ADHD(SD)  \\
        \midrule
        AGE & 11.33 & 0.90 & 11.80 & 0.94  \\
        HOURS OF SLEEP (last night) & 7.33 & 2.26 & 7.67 & 1.80  \\
        VIDEOGAME** & 3.47 & 1.25 & 3.20 & 1.61  \\
        INTERNET** & 3.40 & 1.30 & 3.33 & 1.68  \\
        YEARS OF STUDY & 6.07 & 1.22 & 6.33 & 1.29  \\
        CURRENT SCHOOL GRADE & 5.93 & 1.28 & 5.53 & 1.77  \\
        FAMILIAR INCOME*** & 8.43 & 6.84 & 5.90 & 8,83 \\
        DSM-IN & 2.40 & 1.50 & 7.20 & 1.37  \\
        DSM-IMP+HIP & 2.87 & 1.51 & 4.20 & 2.62  \\
        DSM-TOTAL & 5.27 & 2.40 & 11.40 & 2.75 \\
        IQ & 111.13 & 12.73 & 99.40 & 12.39  \\
        c (task) & 1.80 & 0.01 & 2.00 & 0.00  \\
        q (task) & 1.96 & 0.00 & 2.10 & 0.02  \\
        c (pretask) & 1.68 & 0.01 & 1.90 & 0.01  \\
        q (pretask) & 1.94 & 0.08 & 2.00 & 0.02  \\
        \bottomrule
    \end{tabular}
    }
    
    \label{table}
\end{table*}

The distributions of frequencies reveal a $q$-stretched exponential profile, observing all subjects for each group and condition taken together in the same distribution (global distributions, with n(ADHD) = 15, n(typical) = 15, Fig. ~\ref{fittingall}).  
The refined aspect of the distributions of task condition is due to the larger number of inter-event intervals computed in longer EEG vectors.

Each global distribution was fitted with the function ~\eqref{qstr}. The empirical values found for the task and pretask parameters, were $q$(ADHD) = 2.19, $q$(typical) = 2.34(2.35), $c$(ADHD) = 1.89, and $c$(typical) = 1.71.  The correlations among the four parameters can be seen in Fig. ~\ref{six}:  $c$ shows no significant correlation with the other ones. In parallel, $b$ and $h$ show strong correlation with $q$ ($r$ $>$ 0.80, Pearson Test). The respective linear functions were extracted to analytically set $b$ and $h$ for each group and condition (see methods). Subsequently, $q$ and $c$ were individually estimated by fitting the empirical distribution of each subject with the function ~\eqref{qstr}.

There were no statistical differences between groups concerning confounding variables (age, time using videogame or internet, education, familiar incomes, and hours of sleep at night before the test), indicating no artifact of this kind influencing the results, and consequently that the two groups were paired (see Table ~\ref{table}). The I.Q.(estimated by WISC test) was slightly higher in typical boys than in ADHD ones, which was already pointed out before \cite{Frazier2004, Thaler2013, Mackenzie2016}. However, all subjects of both groups fell within the range of statistically normal I.Q.

\begin{figure*}
\centering
\includegraphics[width=7cm]{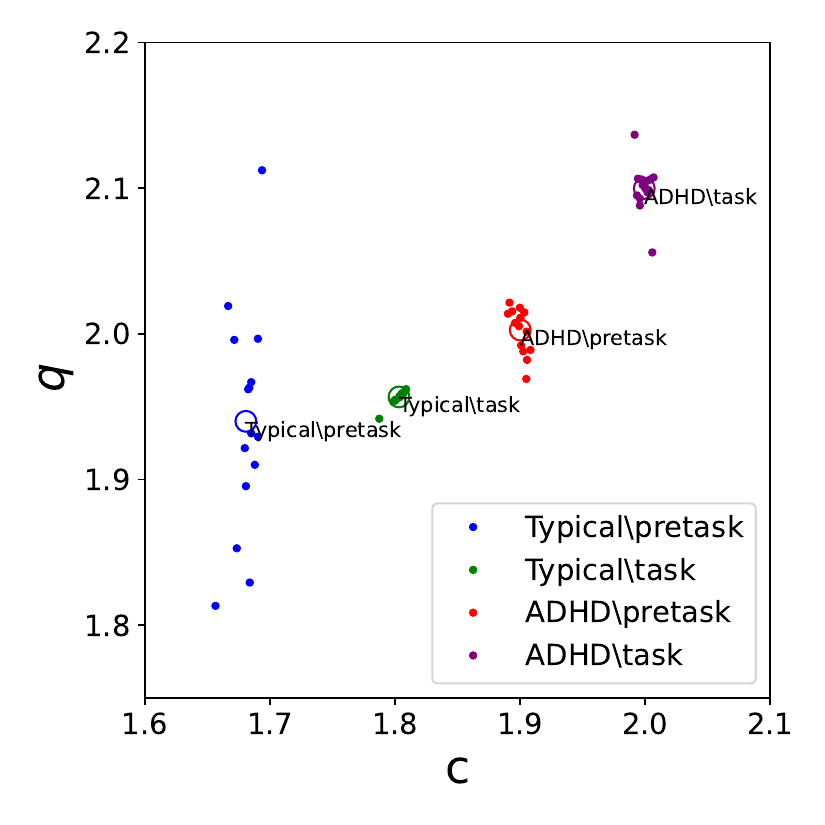}
\includegraphics[width=7cm]{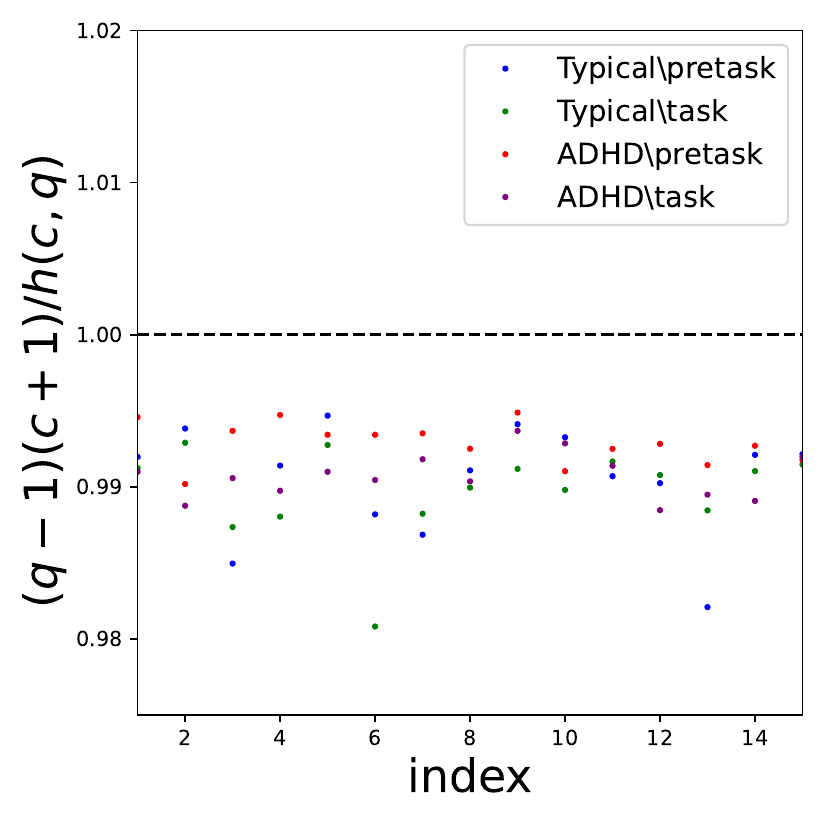}
\centering
\caption{(Left) Linear plot of $q$ versus $c$. The empty circles refer to the mean values of $q$ and $c$ for all cases. (Right) The plot of the condition of normalization of the probability distribution, where $h(q)$ is given by an independent analysis of the four types of electroencephalograms (Typical and ADHD in pretask and task, respectively). }\label{normcond}
\end{figure*}

Unlike the DSM-IV scores for inattention, those for hyperactivity/impulsivity were not different between groups, prevailing the inattentive subtype in the ADHD group (see Table ~\ref{table}).

The averaged parameters $c$ and $q$ from samples were statistically lower for typical subjects compared to ADHD pairs (all $p <$ 0.01, see Table ~\ref{table}), with very low standard deviations. Comparing conditions, only $q$ in ADHD differed between Task and pretask (p = 0.0004, Mann-Whitney U test, see means and std. deviations, Table ~\ref{table}). Fig.~\ref{normcond} shows the space ($c \times q$), where all individual values form well-defined clusters relative to each group and condition. We observed 100 $\%$ accuracy in differentiating typical ADHD in the “task” condition since there was no overlapping regarding the values of ($c \times q$) from these different groups. From all distributions taken together, the global values for $q$ and $c$ differed from averaged ones from individual fittings.

There is no monotonic correlation between $c$ and $q$ and DSM scores for Inattention (Fig.~\ref{DSM}), nor for total scores (result not shown). However, well-defined clusters are observed in scatter plots for both $c$ and $q$(task), which appears to roughly coincide with the DSM cutoff for inattention (more precisely, six criteria are satisfied).
The performance of the subjects in ANT as well as some of their neural correlates have been presented in a previous publication \cite{Abramov2019a}.

\section*{Individual analysis}

\begin{figure*}
\centering
\includegraphics[width=7cm]{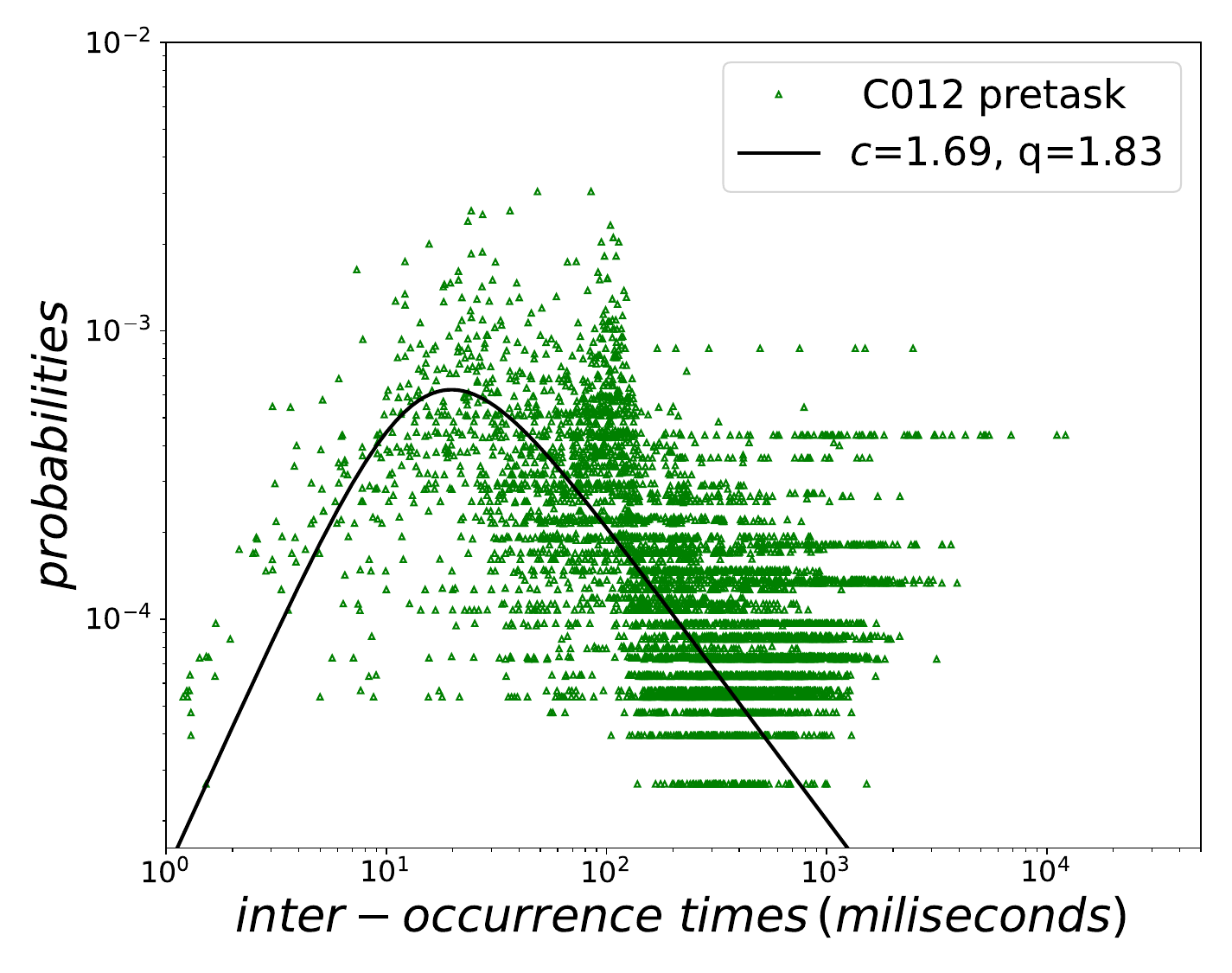}
\includegraphics[width=7cm]{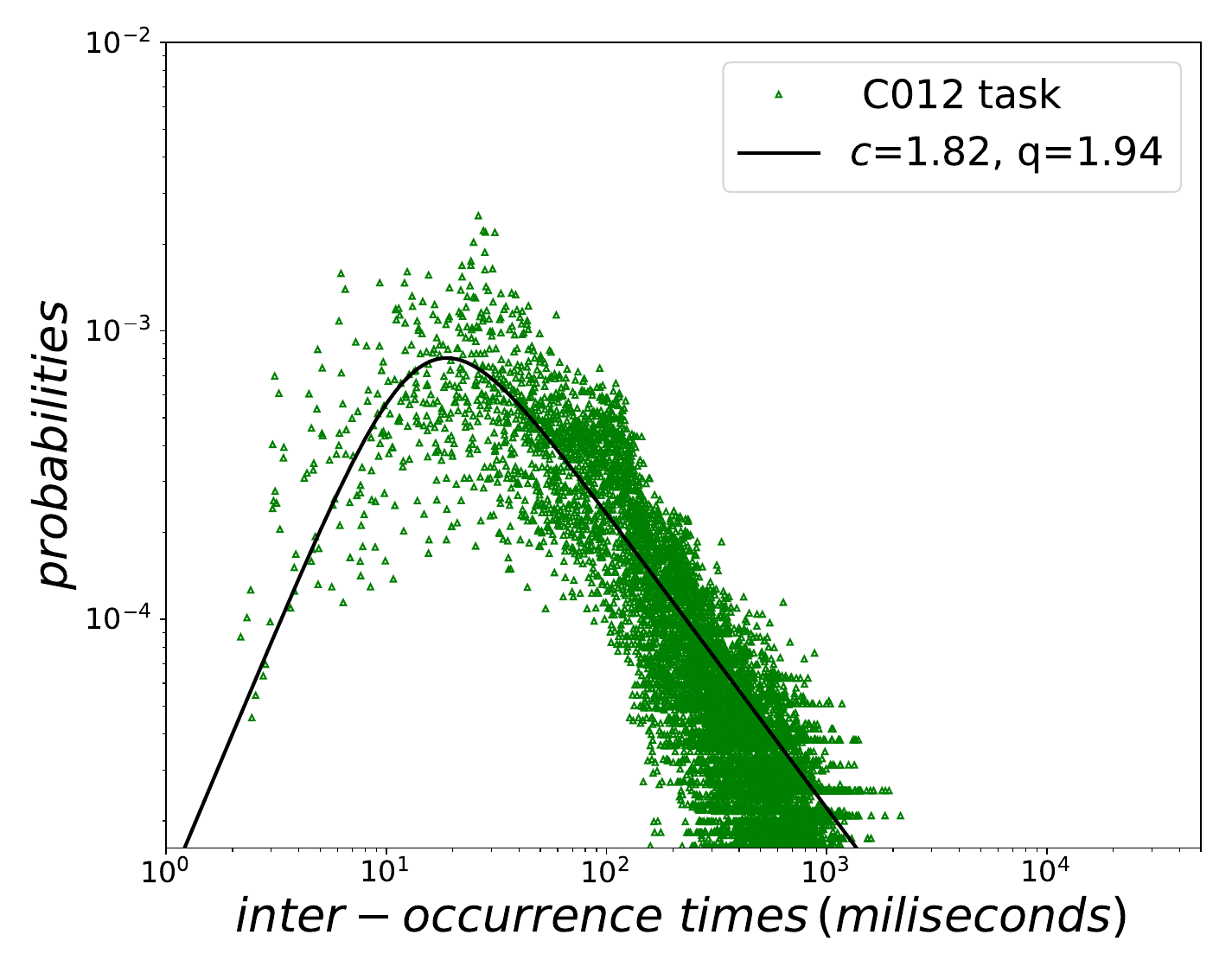}
\includegraphics[width=7cm]{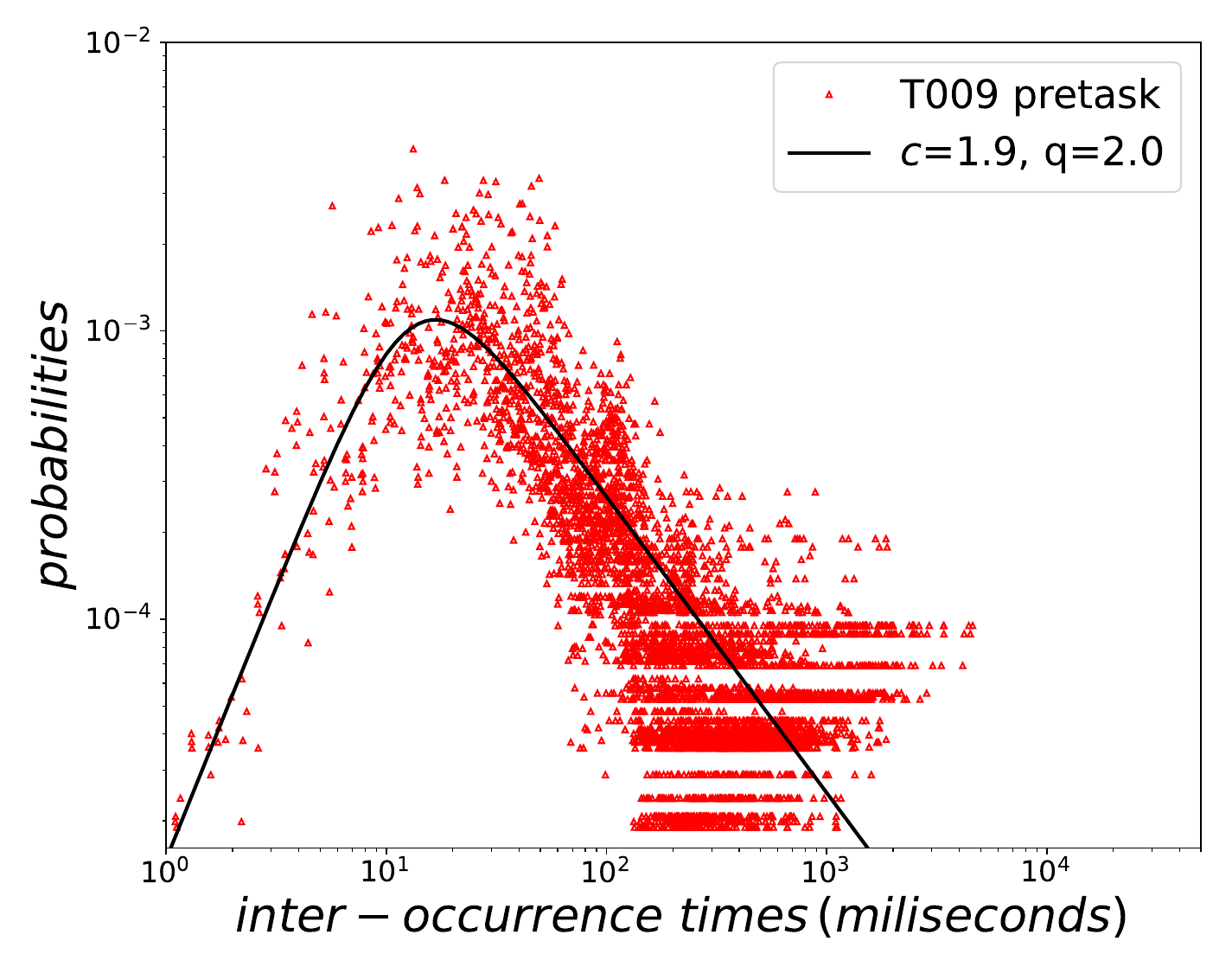}
\includegraphics[width=7cm]{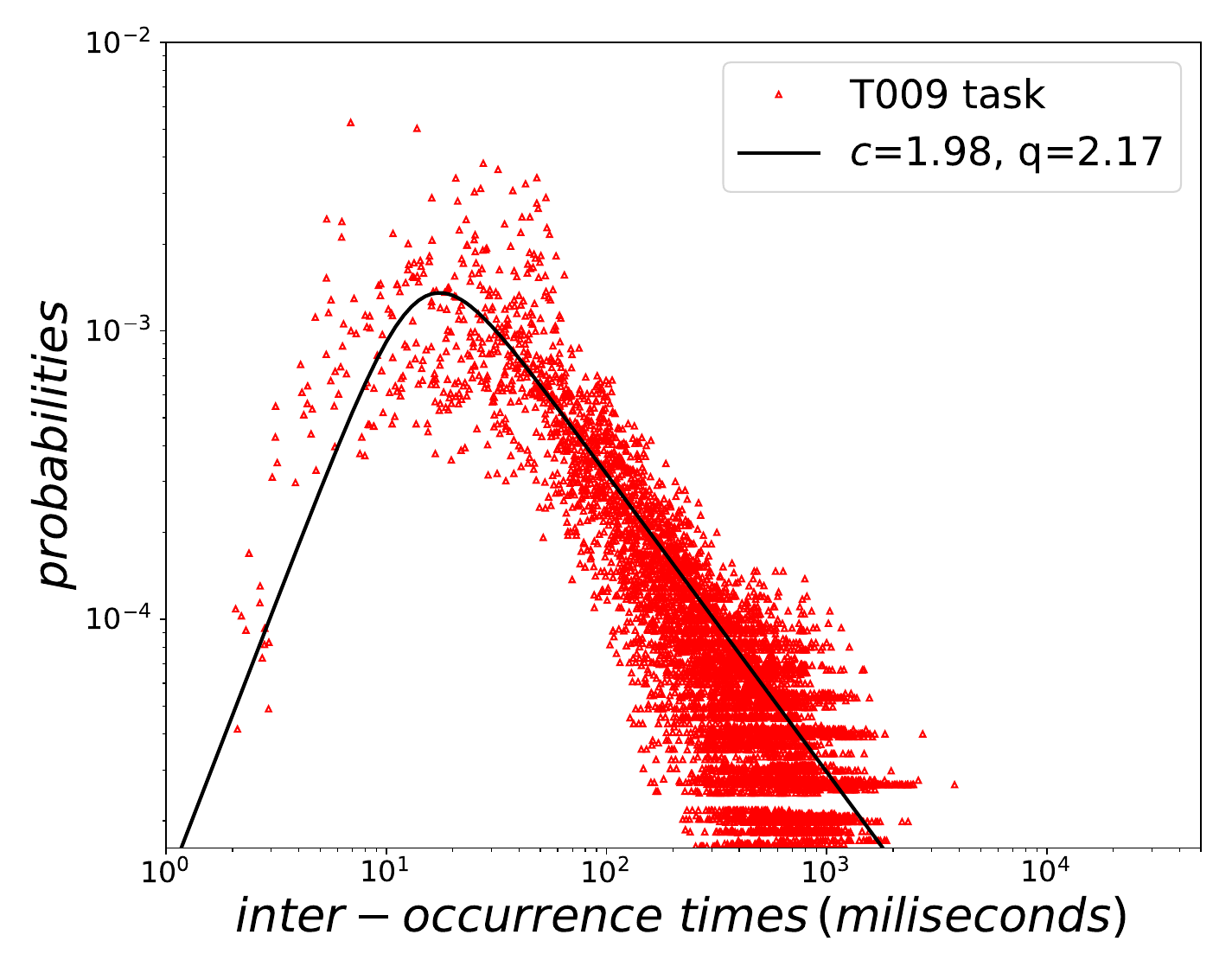}
\centering
\caption{Top: Fittings of the pretask and task EEG's  of the patient C012 (a typical person). Bottom: Fittings of the pretask and task EEG's  of the patient T009 (an ADHD person). }\label{ind}
\end{figure*}

Let us illustrate our methodology through two single examples. In Fig~\ref{ind}, we apply our method to two individuals, one of them being typical (C012) and the other one being ADHD (T009). For each of them, we record the EEG in both pretask and task experimental conditions. We want to determine whether say C012 is typical or ADHD, or even something else. 

We will use the straight lines indicated in Eqs.~\eqref{str} for pretask and task, since these conditions are known in the EEG's of those individuals. 
We then check to verify whether the specific point $(c,q)$ belongs (within some small discrepancy) to one of the clusters in Fig. ~\ref{normcond}. 
For instance, for C012, the values are $(c,q) = (1.69, 1.83)$ and $(c,q) = (1.82, 1.94)$ in pretask and task, respectively. For T009, the values are $(c,q) = (1.9, 2)$ and $(c,q) = (1.98, 2.17)$ in pretask and task conditions, respectively. In contrast, if we use for C012 the wrong straight lines, we obtain $(c,q)=(1.98,2.28)$ (pretask) and $(c,q)=(2,2.2)$ (task), which definitively lie outside all of the clusters in Fig.~\ref{normcond}. If we apply the wrong straight lines to the individual T009 we obtain $(c,q)=(1.7,2.2)$ (pretask) and $(c,q)=(1.78,1.98)$ (task), again outside the clusters. 

Naturally, the performance of the method is expected to become better as the number of patients increases. 

\section*{Discussion}

The traditional analytical-reductionist approach for studying the relationship between the human brain and mind usually comes down to correlating mental processes with the dynamics of neural networks. However, this approach cannot be considered fully adequate, since it does not take into account the complexity of the phenomena that are being compared.

The science of complexity is gaining space every day for studying the cerebral basis of the mind and its disorders \cite{Hernandez2023}. The informativeness in complex systems is non-addictive and non-extensive \cite{Tsallis2009} due to the system's inviolable completeness: the whole is larger than the sum of its parts. Studying the molecular receptor or neuronal cell activity mechanisms, outside their intricate network of correlations, might not be  the most appropriate way to understand brain dynamics related to mental processes. The possibility of using quantitative assessment of NC as a kind of more integral indicator of cerebral functioning seems promising in overcoming the above methodological difficulties in comparing the brain and mind and bringing the solution of these problems to a more adequate level.

\begin{figure*}
\centering
\includegraphics[width=7.4cm]{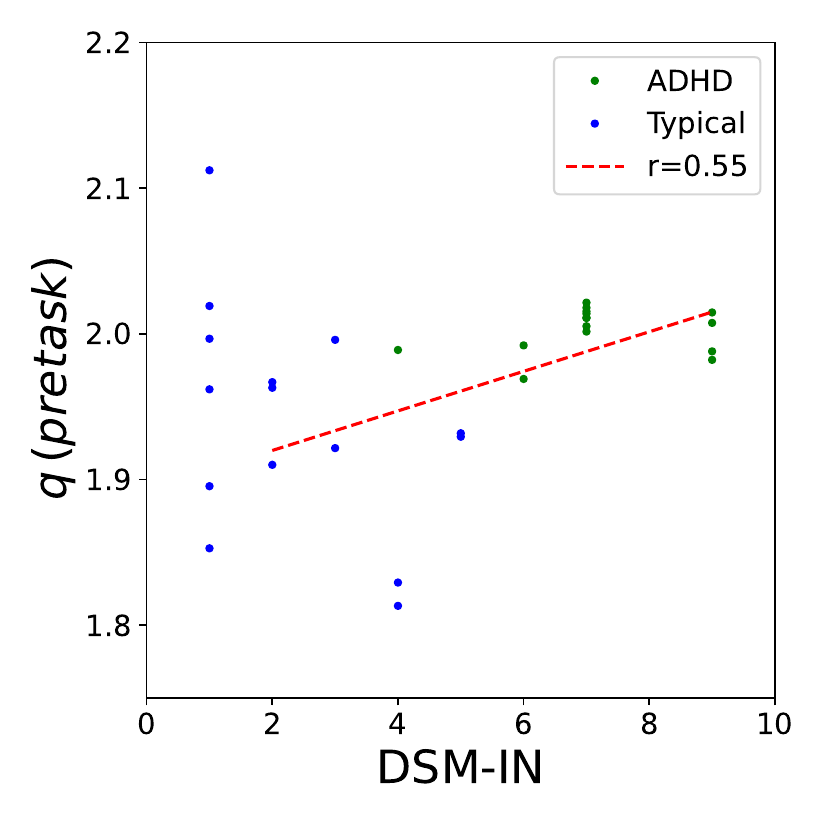}
\includegraphics[width=7.4cm]{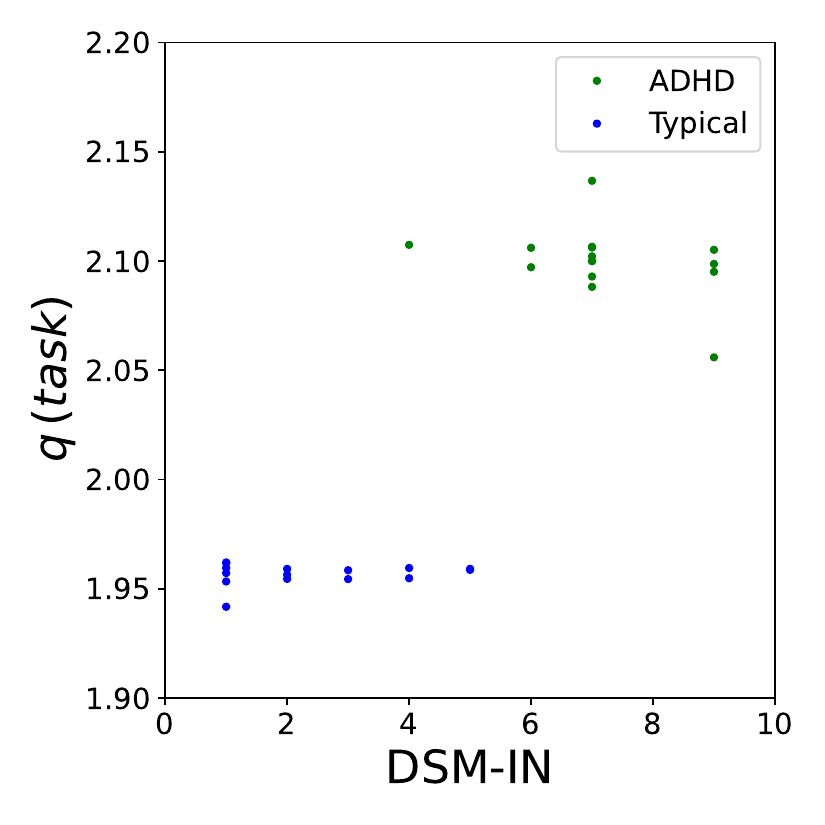}\\
\includegraphics[width=7.6cm]{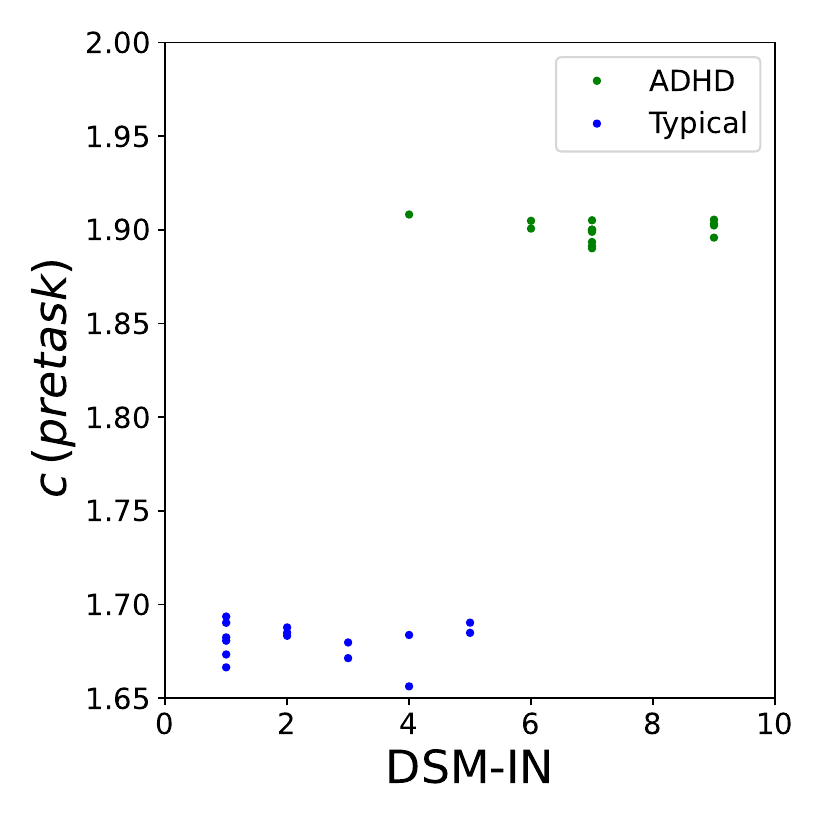}
\includegraphics[width=7.4cm]{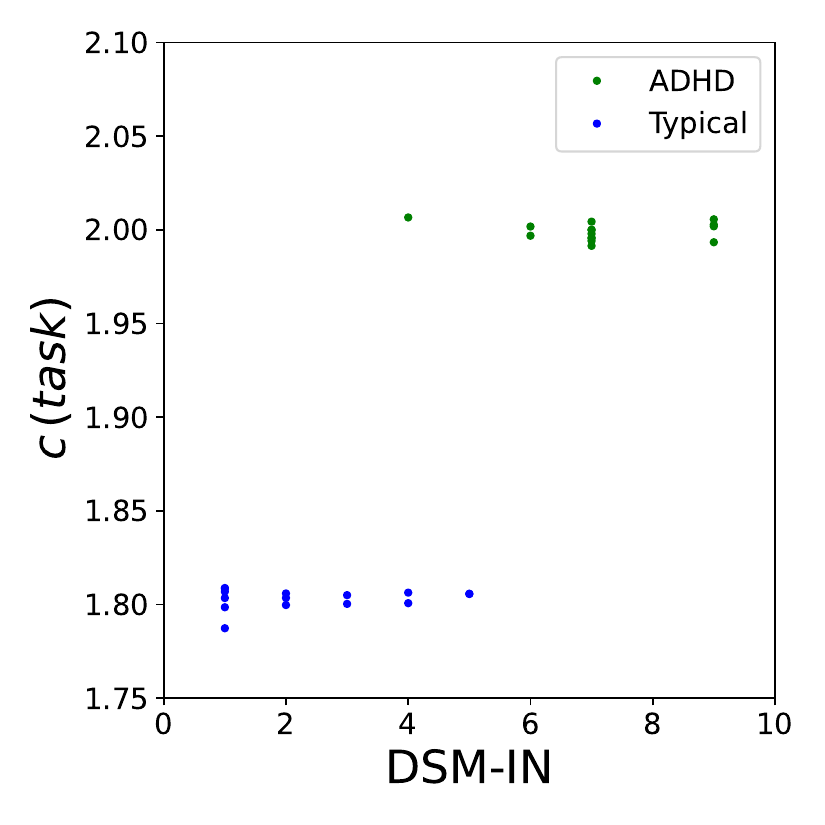}

\caption{ $q$ (pretask), $q$ (task), $c$ (pretask), and  $c$ (task) versus DSM score (top left, top right, bottom left, and bottom right, respectively). Notice that $q$ (pretask) can be roughly fitted with an increasing straight line (dashed red) with $r$(Pearson)=0.55.  }
\label{DSM}
\end{figure*}

 We illustrate the applicability of $q$-stretched exponentials (from NESM) for approaching the EEG complexity\cite{Nosso2023}. 
  Indeed, the parameters $(q,c)$
 appear to accurately discriminate ADHD young boys from their typical pairs. Corroborating other studies \cite{Lauetal2022, Hernandez2023},  which used alternative procedures to infer brain complexity, here the NESM has shown that NC from  EEG of ADHD subjects is higher than that of the typical ones. Let us emphasize that the present approach, based on NESM  \cite{Tsallis1988, Tsallis2009}, enables us to describe the system in terms of complexity using a very simple function involving basic parameters, such as $q$ and $c$.

A hard problem in studying different mental disorders is the reliability of available diagnostic classifications and tools: The DSM (Diagnostic and Statistical Manual of Mental Disorders) is a qualitative/quantitative classifier of mind properties (phenomena, symptoms, and features) to a taxonomy of discrete clusters (diagnostics), designed by expert's perceptions about human diversity \cite{DSMIV, DSM5}. And at least for ADHD, the experts' taxonomy seems to correspond to biological reality. Previously, we had shown high accuracy (nearly 80 percent) among DSM-IV criteria for ADHD and multivariate analysis of the same subjects \cite{Abramov2019b}.

Several previous studies are discordant regarding the neural complexity of typical individuals and those with ADHD, sometimes observing greater or lesser complexity, which complicates drawing an unified  conclusion \cite{Buitelaar2022,Ruiz2023}. However, we observe that such discrepancies are perfectly understandable when considering, first of all, the great variability of factors that are known to influence the structure of the EEG signal, many of which can not be experimentally controlled. For example, the EEG signal varies significantly with age until the end of childhood \cite{Riviello2011}. Comorbidities and treatments undergone, or even the formulation of the diagnosis, which depends on subjective evaluation that is possibly sensitive to cultural biases \cite{Polanczyk2007}, are other confounding factors. Most importantly, we observe that the different brain states evoked during the EEG (whether the eyes are open or closed, or the subject is performing a cognitive task) would have an effect on the measurement of complexity. In this work, we have an homogeneous sampling (boys between 10 and 13 years old) performing a very specific task, which is the ANT. Under these experimental conditions, through the methodology proposed here, we show a possible use of q-statistics to discriminate brain states (pretask and task) and different subjects (typical and with ADHD) through the parameters $c$ and $q$.

However, the design of definitive and universal models for putative brain mechanisms appears to be elusive due to their extremely complex nature. We did not find other studies within which such a high accuracy of agreement between complexity measures and ADHD diagnostics would be achieved~\cite{Fernandez2009, Mohamadi2016, Zarafshan2016, Chen2019}.

Although we have shown that $q$-statistic  can describe NC, it would still be premature to interpret the clinical meaning of the difference in $q$ values found between ADHD and typical subjects. To explain the observed higher complexity in ADHD, we need further studies that we are currently running. In physics, $q$ parameter is related to complexity through long-range correlations, which modify the probability distribution of events as far from the Boltzmann-Gibbs model as the system is complex. EEG studies have found lower beta/alpha or beta/theta rates in ADHD subjects \cite{Barry2003, Loo2012, Barry2013, Markovska2017}, which could contribute to our findings. Whereas gamma oscillations are related to local and bottom-up processing, theta and alpha bands are related to long-range functional connectivity in the brain in top-down (or inner) processing \cite{Stein2000, Lisman2013, Palva2017, Kam2022}. In Fig. \ref{fittingall}, a probability peak on the alpha band shows that this rhythm (between 8-12Hz, which diffusely spreads across the EEG channels) is present in the probability distributions.  


The parameter $c$ is more sensitive to ADHD than $q$, as shown in Fig.~\ref{DSM}. The larger the value of the parameter $c$, the more vertical the left slope of the stretched $q$-exponential function due to the power law $x^c$ prefactor (if $c$=0, the function recovers a  pure $q$-stretched exponential curve). The left slope of the probability distribution is set by the expression of high-frequency EEG bands (beta and perhaps gamma).  EEG gamma frequencies are not easily accessed in clinical examinations. However, there are some studies of the EEG gamma band in relation to psychiatric disorders
 \cite{Newson2019}. Gamma and beta bands are highly correlated to bottom-up cognitive processes \cite{Lisman2013, Palva2017}, and are more expressed in typical than in ADHD subjects \cite{Barry2003, Loo2012, Barry2013, Lazarev2016, Markovska2017}. Indeed, larger $c$ values correspond to lower probability of detecting higher frequencies in the EEG signal since their amplitude is usually below the  threshold that has been used (in preliminary essays, we have observed $c$ = 0 to larger thresholds by statistical formalism in light of the central limit theorem, we advocate in favor of a suitable threshold of one standard deviation). We are currently conducting  another study focusing on the effect of different cognitive states in typical adult subjects upon $c$ and $q$ parameters. 

 As seen in Table \ref{table} and Fig. \ref{fittingall}, the data corresponding to the set of all the subjects lead to $q$ values above the average of the individual ones.  
 This might be due to slight fitting errors related to the nearly straight lines  for $b$ and $h$.

Our first study ~\cite{Nosso2023}  highlighted brain complexity through EEG using q-statistics, where BG statistical mechanics fails.  In this study, we empirically have tested a methodology applying $q$-statistics in the analysis of different mental conditions. We assessed NC through a typical function of $q$-statistics to determine if it could potentially discriminate subtle differences in brain activity between two different human types in different functional brain states (specifically, children with ADHD and typical children, whether performing a cognitive test or not), opening the possibility of using NESM for  diagnostic assessment through the $(c, q)$ map.

Although we currently do not have enough information to establish a more enhanced discussion about the meaning of the EEG complexities and the biological correlates with the $q$ and $c$ parameters, we provide robust evidence for the applicability of $q$-statistics to measure brain complexity and accurately clusterize mental states and conditions. Let us finally emphasize that  investigations involving large numbers of typical and ADHD individuals are highly valuable and welcome. We are presently working along such lines in order to more precisely determine the boundaries of the present four clusters.

\subsection*{Data Archival}

The raw EEG data are disposable upon request.

\section*{Acknowledgment}We would like to thank The Child´s Neurology program of the National Institute of Women, Children, and Adolescents’ Health Fernandes Figueira. This work was supported by Programa de Incentivo a Pesquisa (Research Incentive Program—PIP) of the National Institute Fernandes Figueira (project IFF-008-Fio-13-3-2) as well as by the Brazilian agencies CNPq and Faperj.


\subsection*{References}



%
%

\end{document}